# Clone-Seeker: Effective Code Clone Search Using Annotations

Muhammad Hammad

*Eindhoven University of Technology, Netherlands*

Önder Babur

*Eindhoven University of Technology, Netherlands*

Hamid Abdul Basit

*Prince Sultan University, Saudi Arabia*

Mark van den Brand

*Eindhoven University of Technology, Netherlands*

---

**Abstract**

Source code search plays an important role in software development, e.g. for exploratory development or opportunistic reuse of existing code from a code base. Often, exploration of different implementations with the same functionality is needed for tasks like automated software transplantation, software diversification, and software repair. Code clones, which are syntactically or semantically similar code fragments, are perfect candidates for such tasks. Searching for code clones involves a given search query to retrieve the relevant code fragments. We propose a novel approach called Clone-Seeker that focuses on utilizing clone class features in retrieving code clones. For this purpose, we generate metadata for each code clone in the form of a natural language document. The metadata includes a pre-processed list of identifiers from the code clones augmented with a list of keywords indicating the semantics of the code clone. This keyword list can be extracted from a manually annotated general description of the clone class, or automatically generated from the source code of the entire clone class. This approach helps developers to perform code clone search based on a search query written either as source code terms, or as natural language. In our quantitative evaluation, we show that (1) Clone-Seeker has a higher recall when searching for semantic code clones (i.e., Type-4) in BigCloneBench than the state-of-the-art; and (2) Clone-Seeker can accurately search for relevant code clones by applying natural language queries.

## 1. Introduction

Software plays a central role in society, touching billions of lives on a daily basis. Writing and maintaining source code is a core activity for software developers, who aim to provide reliable and functional software (Allamanis et al., 2015). One of the challenges a software developer faces when writing new code is to find out how to implement a certain functionality (e.g., how to implement quick sort) (Kim et al., 2018). The implementation of such functionality might already be realized by other developers and can be reused rather than written from scratch. Over the years, a huge number of open source and industrial software systems have been developed and the source code of these systems is typically stored in source code repositories such as GitHub. This source code can be treated as an important reusable asset for developers (Lv et al., 2015).

Many software development and maintenance tasks rely on effective code search (Sadowski et al., 2015) to find the source code related to a specific functionality. In modern software development, developers often refer to web search engines in order to search for code examples from large amounts of online resources such as GitHub, online tutorials, technology blogs, API documents, social media posts, etc (Keivanloo et al., 2014; Zhong et al., 2009). Indeed, code search is an integral part of software development; developers spend up to 19% of their development time on code search (Ko et al., 2005). Similarly, studies have even revealed that more than 60% of developers search for source code examples every day (Hoffmann et al., 2007; Stolee et al., 2014).

Source code examples or code fragments can help developers understand how others addressed the similar problems (Ko et al., 2006; Xie and Pei, 2006; MP, 2009; Bajracharya and Lopes, 2012; Martie et al., 2017; Xia et al., 2017) and can serve as a basis for writing new programs (Johnson, 1992; Kim et al., 2010; Buse and Weimer, 2012). These code examples can accelerate the development process (Mandelin et al., 2005) and increase the product quality (Marri et al., 2009). A working code example can be considered for both learning and pragmatic reuse. Such working code examples can spawn a wide range of applications, varying from API usage (e.g., how to use the JFreeChart library[1] to display a chart in Java) to basic algorithmic problems (e.g., how to implement quick sort). An ideal working code example should be concise, complete, self-contained, and easy to understand and reuse. Code clones can be considered as the ideal code examples as they are more stable and possess less risk than new development (Kapser and Godfrey, 2008; Hammad et al., 2020a).

---

[1] https://www.jfree.org/jfreechart/

Code clones are usually categorized as Type-1, Type-2, Type-3 and Type-4 (Roy et al., 2009; Hammad et al., 2020b) clones, based on their level of similarity with each other. Type-1 clones are same code fragments, except for dissimilarities in comments, layout, and whitespace. Type-2 clones refer to same code fragments, except for dissimilarities in literal values and identifier names, in addition to Type-1 clone dissimilarities. Type-3 clones are syntactically similar code fragments with differences in their statements. Such fragments can contain additions, modifications and/or removals to their statements with respect to each other, in addition to the changes allowed for Type-1/-2 clones. Type-4 clones are also known as semantic clones, which have similar functional behavior, even if the syntax of the code is different. Developers often need to search for these code clones to improve their code (Kim et al., 2018) for addressing software engineering challenges such as software diversification (Baudry et al., 2014), software repair (Ke et al., 2015), and even automated software transplantation (Barr et al., 2015). According to Kapser and Godfrey (Kapser and Godfrey, 2008), code clones are helpful in exploratory development, where one wishes to rapidly develop a new feature using a clone-and-own approach, and not necessarily unify (refactor, parameterize, etc. ) the existing clones. Hence, such a cloning approach allows flexibility and increased productivity in an opportunistic programming scenario.

However, searching for code clones (or code fragments in general) is challenging, because there is only a small chance (10-15%) that developers would guess the exact words used in the code (Furnas et al., 1987) and use them in their query. Similarly, source code is structured non-linearly; this makes it difficult to be read in a linear fashion like normal text, and be searched effectively. There are several other factors which depends on the effectiveness of code search such as quantity of data, quality of the indices (Pathak et al., 2000; Kim et al., 2018), search technique, query, and metadata (also known as *natural language document*). Often it is difficult for a developer to formulate an accurate query to express what is really in her/his mind, especially when the maintainer and the original developer are not the same person. When a query performs poorly, it has to be reformulated. But the words used in a query may be different from those that have similar semantics in the source code, i.e., the synonyms, which will affect the accuracy of code search results. For example, in Java, the programming concept "array" does not match with its syntactic representation of "[ ]". Code search engines can more effectively assist developers over the search query, if such semantic mappings exist. Similarly, there are two types of code search engines. One is known as code-to-code search engine, which accepts code fragments from users, and recommends syntactically or semantically similar code fragments found in a target code base (Kim et al., 2018). The other one is known as natural-language-to-code search engine, which accepts natural language terms and recommends code fragments from a target code base that closely match those terms. Implementing any of these search engines can be challenging, and recently a number of techniques have been proposed to address the weaknesses of existing code search techniques (Hill et al., 2011; Yang and Tan, 2012; Haiduc et al., 2013).

We found that the accuracy of existing code search tools are often unsatisfactory in retrieving Type-4 (semantic) clones, with the recall for Type-4 clones reported by the state-of-the-art clone detectors and clone search approaches is as low as 17% (Ragkhitwetsagul and Krinke, 2019). In this work, we tackle the problem of effective code-to-code search, particularly for Type-4 clone methods. We evaluate the effectiveness of our technique by using natural language queries. We propose a number of different approaches to represent metadata for each clone method to enable accurate and efficient retrieval. Our experiments show that the accurate representation of clone methods in terms of metadata increases the effectiveness of code search. Specifically, we have made the following contributions:

1. We present a novel approach called Clone-Seeker, which builds a natural language document of each clone method by combining important keywords of each clone method with keywords extracted from a general description of the clone class, annotated either manually or automatically.

2. Our approach can successfully assist developers in performing code-to-code search and natural language query search.

3. We have quantitatively evaluated our approach using the BigCloneBench dataset for code-to-code search and Stack Overflow for natural language queries in terms of the quality of the search results.

## 2. Related Work

To the best of our knowledge, no previous technique has explored the effect of utilizing clone class features in searching for clone methods. However, there are several approaches in the literature focusing on different aspects of effective code search such as query refinement, quality of indices and search technique. We discuss these approaches in this section.

Several techniques for code search have been devised, which focus on query refinement and query expansion. For example, Hill et al. (2011) reformulate queries with natural language phrasal representations of method signatures. Haiduc et al. (2013) proposed to reformulate queries based on machine learning. They trained a machine learning model that automatically recommends a reformulation strategy based on the query properties. Lu et al. (2015) proposed to extend a query with synonyms generated from WordNet. However, Sridhara et al. (2008) observed that automatically expanding a query with inappropriate synonyms may produce even worse results than not expanding the query.

There is also a substantial volume of work that takes into account code characteristics for code search. For example, McMillan et al. (2011) proposed Portfolio, a code search engine that combines keyword matching with PageRank to return a chain of functions. Lv et al. (2015) proposed CodeHow, a code search tool that incorporates an extended Boolean model and API matching. Ponzanelli et al. (2014) introduced an approach that automatically retrieves pertinent discussions from Stack Overflow, given a context in the IDE. Li et al. (2016) present RACS, a code search framework for JavaScript that considers relationships (e.g., sequencing, condition, and callback relationships) among the invoked API methods.

Our approach is quite similar to the idea of Software Bertillonage (Davies et al., 2011), which is a signature based matching technique. It focuses on the reduction of search space, when trying to locate a software entity within bytecode. In other related work, researchers proposed representing code snippet in the form of metadata or natural language document or signature and applied different search techniques such as information retrieval, neural networks, and joint models. For example, Sachdev et al. (2018) investigated the use of natural language processing and information retrieval techniques to carry out natural language search directly over source code. They represented code snippet in the form of natural language document to make the retrieval of code effective. Kim et al. (2018) proposed FACOY, which is an effective code-to-code search engine for finding semantically similar code fragments in large code bases. FACOY takes a code snippet as the input query and retrieves semantically similar code snippets from the corpus. Given a code query, it first searches in a Stack Overflow dataset to find natural language descriptions of the code, and then finds related posts and similar code. Sourcerer (Bajracharya et al., 2006) is a framework for performing code search over open–source projects available on the Internet. Sourcerer works by extracting keywords and fine–grained structural features from source code, and searching for similar code using the text search engine Apache Lucene[2]. Ragkhitwetsagul and Krinke (2019) incorporate a multi-representation technique, corresponding to four clone types, to represent an indexed corpus of code. They improve the query quality by leveraging the knowledge of token frequency in the codebase, and finally re-rank the searched candidate code based on the TF-IDF weighting method. Zhou et al. (2019) proposed Lancer, a context-aware code to-code recommending tool. Lancer uses a Library-Sensitive Language Model and a BERT model to recommend relevant code samples in real-time based on the incomplete code. There are various search methodologies (Gu et al., 2018; Husain et al., 2019; Allamanis et al., 2015) which jointly embed code snippets and natural language descriptions into a high-dimensional vector space, in such a way that code snippet and its corresponding description have a similar vector representation.

## 3. Our Methodology for Code Clone Search

In this section, we outline our methodology for Clone-Seeker, which focuses on utilizing clone class features in retrieving code clones, given a search query. The effectiveness of code clone search relies on multiple factors such as search technique, quality of the search query, and its relationship to the text contained in the software artifacts (metadata, i.e. natural language document). We believe that utilizing clone class features should also be considered as an important factor, which can help in performing effective code clone search. For this purpose, we first apply pre-processing steps to extract identifiers from each clone method. Afterwards, we present two ways to annotate clone classes with keywords: manual and automatic. Finally, we augment the annotated keywords of clone classes with clone method identifiers to build a natural language document of each clone method. Figure 1 displays a pictorial representation of our methodology. The top portion displays how our search corpus is built from a dataset, whereas the bottom portion displays how the search results are retrieved from Clone-Seeker. We elaborate the details of our approach in the following sections.

---

[2]`https://lucene.apache.org/`

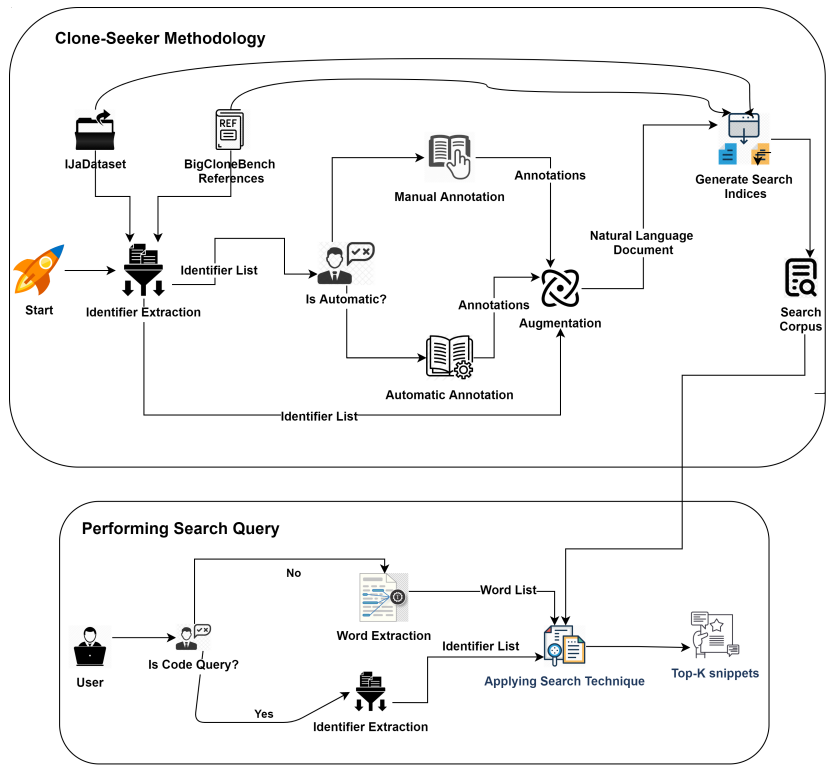

Figure 1: Clone-Seeker Methodology and Applying Search Query

### 3.1. Dataset Selection

We build our search corpus from IJaDataset clone methods referenced by BigCloneBench (Sva-jlenko et al., 2014; Svajlenko and Roy, 2016). BigCloneBench is the largest clone benchmark dataset, with over 8 million manually validated clone method pairs in IJaDataset 2.0[3]. IJaDataset, in turn, is a large Java repository with 2.3 million source files (365 MLOC) from 25,000 open-source projects. BigCloneBench contains references to both syntactic and semantic method clones, and keeps the references of starting and ending lines for those clones. In forming this benchmark, the authors used pattern-based heuristics for identifying the methods that potentially implement a given common functionality. These methods were manually annotated as true or false positives of the given functionality by the judges. All true positives of a functionality were grouped as a clone class, such that a clone class of size $n$ contains $\frac{n(n-1)}{2}$ clone pairs. Currently, BigCloneBench contains clones corresponding to 43 distinct functionalities (i.e. clone classes).

### 3.2. Identifier Extraction

Identifiers are an important source of domain information and can often serve as a starting point in many program comprehension tasks (Fritz et al., 2014; Lu et al., 2015). We follow similar pre-processing steps to extract identifiers from each clone method as defined by Lu et al. (2015). First, we extract starting and ending line references of a total of 14,922 true positive clone methods (*Extraction*). Next, we trace them in the IJaDataset files, by following their references from the BigCloneBench dataset, and put them in our search corpus list (*Tracing*). Afterwards, we normalize each clone method code by removing the Java reserved keywords, constant values, whitespaces, extra lines, comments, as well as perform tokenization (*Normalization*). We use the Javalang[4] Python library for tokenization, which contains a lexer and parser for the Java 8 programming language. We do not use comments as they have been reported to be non-reliable and inconsistent source for extracting natural language document (Yan et al., 2020; Ragkhitwetsagul et al., 2019). Similarly, software projects can be poorly documented. We are unable to know the extent to which each comment accurately describes its associated clone method. For example, a number of comments can be outdated with regard to the code that they describe. Moreover, we found some comments of clone methods in IJaDataset written in other languages, whereas our methodology focuses only on

---

English queries. After tokenization, we perform snake case and camel case splitting of identifiers to separate words(*Splitting*). We also eliminate single characters (*Single Character Elimination*) and apply stemming(*Stemming*). This finally produces a flat list of keywords representing each clone method.

### 3.3. Annotating Code Clones

Data annotation is a process of labeling the data available in various formats like text, video or images (Pustejovsky and Stubbs, 2012). It is expected to play a major role in enhancing product recommendations, relevant search engine results, computer vision, speech recognition, chatbots, and more. We apply annotation techniques to best describe the core functionality of each clone class in the BigCloneBench dataset. The purpose of annotating clone classes is to assist in retrieving clone methods, when a code-to-code search or a natural language query is applied. Annotation can be performed manually or automatically, the details of which are described as follows.

*Method 1: Manual Annotation.* Manual annotation is a process of labeling or annotating any data by humans. The approach is popular owing to its benefits such as accuracy, high level of integrity, need for minimal administration of data annotation efforts, and a higher chance of discovering intriguing insights pertaining to the data as compared to automatic annotation techniques, which can be later integrated into an algorithm. However, manual annotation is more expensive and time-consuming. Nowadays, there are many key players offering exclusive products and services in the market[5], to manually annotate a huge bulk of data.

BigCloneBench dataset is a manually created and validated dataset, which also contains annotations, i.e. the description of each clone class in natural language terms. To use in this paper, we manually changed the description slightly by removing bullet points, correcting spelling mistakes, and making the sentence structure more readable. Table 4 displays the resulting list of clone class descriptions used in our experiments. Afterwards, we performed several pre-processing steps such as removing spaces, single characters, and stop-words, and performed stemming in order to get a flat list of words (*Word Extraction*) per clone class.

*Method 2: Automatic Annotation.* The second approach that we adopt to annotate clone classes involves automatically extracting keywords. Keywords are a subset of words or phrases from a document that can describe concepts or topics covered in the document (Hulth, 2003; Zhang, 2008). They are commonly used to annotate articles or other documents, and are essential for the categorization and fast retrieval of such items in digital libraries (Çano and Bojar, 2019).

---

[5] `https://aimultiple.com/data-annotation-service`

Table 1: Manual annotation of Clone Classes

| Id | Description |
|---|---|
| 2 | Download file from http link by using Web |
| 3 | Generate secure hash in binary format and convert it into a string representation |
| 4 | Copy file from source to destination |
| 5 | Retrieve a zip archive (from disk, internet, etc). Open it, and iterate through its entries. Decompress the files to disk, by preserving their directory structure. |
| 6 | Connect to the FTP server by log in with user name and password |
| 7 | Sorts an array of values using bubble sort |
| 8 | Setup SGV by creating ScrollingGraphicalViewer object and inject model into it |
| 9 | Setup SGV event handler by creating ScrollingGraphicalViewer object and add listener |
| 10 | Execute a database update and rollback perhaps conditionally |
| 11 | Initialize Java eclipse project by creating it in a workspace. Then, set the project's nature to Java (may also set other natures, like plugin). Afterwards, set its class path. Finally, set its output path as bin directory or folder. |
| 12 | Takes an integer and returns its prime factors |
| 13 | Shuffle array by placing elements randomly such as by using Fisher-Yates algorithm |
| 14 | Finds the position of a specified input value (search key) within an array by key value using the binary search algorithm. True positive may use any linear data-structure (array, list, etc). Must use the binary search algorithm. Must return the index or an error code. |
| 15 | Load custom font by using URL or file etc. Then, create font and register it with graphics environment. |
| 17 | Create encryption key files by first generating a pair of encryption keys such as public and private. Then, these keys are written to file or files in some format. |
| 18 | Takes a sound source and plays it |
| 19 | Take screenshot. Save screenshot to file. |
| 20 | Calculates either the ith fibonacci sequence, or the fibonacci sequence up to (or including) the ith term. |
| 21 | Build a message such as MessageBuilder and send it over XMPP service. |
| 22 | Takes some data, encrypts it, and writes it to a file |
| 23 | Resizes (shrink or expand) an array |
| 24 | Open URL in system default web browser. |
| 25 | Open file in desktop application by firstly check if desktop AWT is supported on the platform (isDesktopSupported). Then, open the file by using default desktop application. |
| 26 | Find Greatest Common Denominator (GCD) |
| 27 | Uses reflection to access a method and calls or invokes that method |
| 28 | Parse XML to DOM by first creating or retrieving DocumentBuilderFactory. Then, configure that factory and use it to create new document builder. Finally, use that document builder to parse the XML data and return or use the DOM. |
| 29 | Convert date string format by parsing source date string and format it by using a destination format. |
| 30 | Create zip archieve containing one or more files |
| 31 | File selection dialog. Use a file chooser to select one or more files or directories. |
| 32 | Send email using Java mail. Receives a session (either creates session in snippet, or its created elsewhere). Creates the mime message and configures it with at least some standard data. Sends the email. |
| 33 | Receives some file. Produces a CRC32 checksum of the file. |
| 34 | Execute external process and do something with the output (stdout or sterr or both). |
| 35 | Instantiate using reflection. Get class object. Use reflection to find the constructor. Use the constructor to instantiate the object. |
| 36 | Connect to database. Create a database connection. |
| 37 | Load file into byte array. The entire file is loaded into a single byte array. |
| 38 | Get the MAC address of a network device. Convert byte MAC address into standard format. |
| 39 | Delete a folder, and all the files and folders it contains, recursively down the file heirarchy. |
| 40 | Parse the elements of a csv file. Where each line is an entry, and the elements of an entry are separated by some delimiter on that line. |
| 41 | Transpose a matrix (2d array). |
| 42 | Use a regular expression to extract matches of a pattern in text. |
| 43 | Copy directory and its contents. |
| 44 | Test if a string is a palindrome. |
| 45 | Writes a pdf file. |

Automatic keyword extraction (also known as keyword detection or keyword analysis) is the process of selecting words and phrases from a text document that can best project the core sentiment of the document without any human intervention depending on the model (Zhang, 2008; Hulth, 2003). It is a text analysis technique that automatically extracts the most used and most important (with respect to certain criteria, to be elaborated later in this section) words and expressions from text. It helps to summarize the content of texts, recognize the main topics discussed, and automatically create a compressed version of a text that provides useful information for the users.

Keyword extraction simplifies the task of finding relevant words and phrases within unstructured text. There are various methodologies to extract keywords such as word frequency, word degree, TF-IDF and RAKE (Nasar et al., 2019). We follow a simple approach known as word frequency or naive counting (Iker, 1974; Baron et al., 2009), in which we identify the list of words that repeat the most within a set of natural language documents of clone methods. This can be useful for identifying recurrent terms in a set of clone methods belonging to a clone class, and eventually finding out the most common words in the whole clone class.

### 3.4. Augmentation of Identifiers With Annotated Keywords

We augment the list of words from the identifier extraction process (Section 3.2) with the ones extracted from the clone class annotations (Section 3.3) and build a natural language document (i.e. metadata) for each clone method. We believe that the augmentation will help in retrieving clone methods effectively when a search query is applied. For further illustration, we explain how the natural language document of a "Copy File" clone method has been built in Table 2, by merging the identifier keywords with the annotation words. Annotated words are highlighted with blue color.

### 3.5. Formulating a Search Query

There are two ways to formulate a search query; one is to enter some code fragment as a query, and the other is to write keywords in natural language to find relevant code fragments. Our approach can help in retrieving clone methods in both these forms. For the first case, i.e. code-to-code search, we perform the pre-processing steps on the search query in the form of a code method, as mentioned in *Identifier Extraction* (see Section 3.2). In the second case, we perform the pre-processing steps for *Word Extraction*, in order to get a flat list of words.

Table 2: Building the natural language document of the "Copy File" clone method

| | |
|---|---|
| **Clone Method** | ```java
public static void copyFile(File src, File dest) throws IOException {
    FileInputStream fis = new FileInputStream(src);
    FileOutputStream fos = new FileOutputStream(dest);
    java.nio.channels.FileChannel channelSrc = fis.getChannel();
    java.nio.channels.FileChannel channelDest = fos.getChannel();
    channelSrc.transferTo(0, channelSrc.size(), channelDest);
    fis.close();
    fos.close();
}
``` |
| **Identifier Extraction** | copi file src dest io except input stream fis output fos java nio channel get transfer to size close |
| **Manual Annotation** | Copy a file from source to destination |
| **Manual Annotation (after pre-processing)** | copi file sourc destin |
| **Natural Language Document** | copi file sourc destin src dest io except input stream fis output fos java nio channel get transfer to size close |

### 3.6. Search Methodology

Information Retrieval (IR) techniques have been successfully applied to address various software engineering tasks, including concept/feature/concern location, impact analysis, code retrieval and reuse, bug triage, refactoring and restructuring, reverse engineering, and defect prediction(Marcus and Antoniol, 2012). IR techniques, in general, are used to discover the significant documents in a large collection of documents, which match the user's query. They mainly aim for identifying the information relevant to the user requirements in a given scenario. An IR-based code retrieval method in particular usually extracts a set of keywords from a query and then searches for the keywords in code repositories (Nie et al., 2016).

We apply an IR technique to retrieve the top-10 results from the search corpus that match the natural language query. The selected IR technique is based on Term Frequency-Inverse Document Frequency (TF-IDF) word embeddings for retrieving the clone methods most similar to the query (Dillon, 1983). TF-IDF is a technique often used in IR and text mining. According to a survey in 2015, 70% of text-based recommendation systems for digital libraries use TF–IDF (Beel et al., 2016). TF-IDF uses a weighting scheme which assigns each term in a document a weight corresponding to its term frequency and inverse document frequency. In our context, TF-IDF looks at the term overlap, i.e. the number of shared tokens between the two clone methods in question (and also how important/significant those tokens are in the clone methods). We use TF-IDF with unigrams as terms to transform clone methods into numeric vectors. These vectors can in turn easily be compared by quickly calculating their cosine similarities. If a term appears frequently in a clone method's natural language document, that term is likely important in that method. The frequency of a term is simply the number of times that a term appears in a clone method. However,



if a term appears frequently in many clone methods' natural language document, that term is likely less important generally. To factor this, we use the IDF measure. IDF is the logarithmically-scaled fraction of clone methods in the corpus in which the term appears. The terms with higher weight scores (high TF *and* IDF) are considered to be more important. We first transform clone methods existing in the search corpus and the pre-processed query into TF-IDF vectors using the formula in Equation 2.

$$TF - IDF(i, j) = (1 + \log(TF(i, j)). \log(\frac{J}{DF(i)})$$

(1)

where *TF (i, j)* is the count of occurrences of feature $i$ in clone method $j$, and *DF (i)* is the number of clone methods in which feature $i$ exists. $J$ is the total number of clone methods. For the retrieval, we generate a normalized TF-IDF sparse vector from a given query, and then take its dot product with the feature matrix. Given that all vectors are normalized, the result yields the cosine similarity between the query vector of the query and of every clone method. Afterwards, we return the list of all the clone methods, ranked by their cosine similarities to the query vector.

## 4. Empirical Evaluation

We present empirical results in this section to validate the effectiveness of our approach. We evaluate Clone-Seeker by formulating the search query in two ways: by source code and by natural language terms. For the first case, we compare our results for the code-to-code search approach with previous experimental studies across various clone types including semantic clones. For the second case, we evaluate the effectiveness of our methodology for code clones when a natural language query is applied.

### 4.1. Code-to-Code Search Using Code Clones

In this section, we demonstrate the effectiveness of Clone-Seeker by applying code as a search query (code-to-code search).

*Experiment Design.* We evaluate Clone-Seeker against an existing benchmark in a scenario for searching by code. We compare recall of our search approach with state-of-the-art clone detector and search approaches presented by Kim et al. (2018); Ragkhitwetsagul and Krinke (2019).

In BigCloneBench, clone pairs are assigned a type based on the criteria in (Sajnani et al., 2016). Type-1 and Type-2 clone pairs are classified according to the classical definitions given in Kim



et al. (2018); Ragkhitwetsagul and Krinke (2019). Moreover, Type-3 and Type-4 clones are divided into four sub-categories according to their syntactical similarity: Very Strongly Type 3 (VST3), Strongly Type 3 (ST3), Moderately Type 3 (MT3), and Weakly Type 3/Type 4 (WT3/4). Each clone pair (unless it is Type 1 or 2) is identified as one of the four types if its similarity score falls into a specific range; VST3: [90%, 100%), ST3: [70%, 90%), MT3: [50%, 70%), and WT3/4: [0%, 50%).

Given that BigCloneBench consists of clone classes each having a set of clone methods (i.e. references to Java methods in IJaDataSet), we process each individual clone method to be used in a code-to-code query. For each clone method $c_{input}$, we first extract the identifiers from the source code following the steps mentioned in Section 3.2. Using this flat list of identifiers as the query to Clone-Seeker, we search for the most similar natural language documents as developed in Section 3.4. These natural language documents links to resulting code methods ($c_{res}^{k..l}$) in IJaDataset. Since the most similar code methods are already labelled and referenced in BigCloneBench ($c_{ref}^{m..n}$) in the corresponding clone class, we check whether the pairs ($c_{input}$,$c_{res}^i$) coincide with any ($c_{input}$,$c_{ref}^j$). We compute the recall of Clone-Seeker following the definition proposed in the original benchmark of BigCloneBench (Svajlenko et al., 2014).

$$Recall = \frac{D \bigcap B_{tc}}{B_{tc}} \tag{2}$$

where $B_{tc}$ is the set of all true clone pairs in BigCloneBench, and D is the set of clone pairs found by Clone-Seeker. We evaluate three different approaches as the annotation strategy: (1) Manual Annotation and (2) Automatic Annotation, as described in Section 3.3, as well as (3) Baseline. In Baseline, we do not provide any annotation, and aim for assessing the impact of adding annotations on the accuracy of code-to-code search. We compare the performance of Clone-Seeker with the other approaches reported in Kim et al. (2018); Ragkhitwetsagul and Krinke (2019). For fairness in comparison, we choose the same configuration of retrieving the top-900 natural language documents, which are mapped to their associated clone methods. In order to annotate each clone class automatically with a set of words, we can retrieve the top-k most recurrent words list by applying different threshold values for $k$ such as 5, 10, 15, 20, and 25. In our preliminary experimentation on a random subset of 100 queries, we observed that k=10 gave us the best result in terms of recall. Therefore we choose the top-10 most recurrent keywords for our experiments in this paper when using automatic annotation. We then proceed to calculate the recall after performing



the augmentation step (see Section 3.4).

*Results.* Table 3 depicts the recall scores for our approach with different annotation strategies: Clone-Seeker(Baseline), Clone-Seeker(Manual), and Clone-Seeker(Automatic). Recall scores are summarized per clone type with the categories introduced above. Since for Clone-Seeker we are reproducing the experiments performed in Sajnani et al. (2016); Kim et al. (2018); Ragkhitwetsagul and Krinke (2019), we directly report in the same table all the results that the authors have obtained on the benchmark for their own approaches such as Siamese, FaCoY and state-of-the-art code clone detectors including NiCaD (Cordy and Roy, 2011), iClones (Göde and Koschke, 2009), SourcererCC (Sajnani et al., 2016), CCFinderX(Kamiya et al., 2002), and Deckard(Jiang et al., 2007).

Our goal, as outlined in Section 1, is to build Clone-Seeker as a code-to-code search engine capable of finding Type-4 clones. Nevertheless, for a comprehensive evaluation of the added value of the strategies implemented in our approach, we provide the comparative results of recall values across all clone types. Overall, we notice that the recall performance of Clone-Seeker(Automatic) is better than Clone-Seeker(Baseline) and Clone-Seeker(Manual) across all clone types. The three Clone-Seeker variants produce the highest recall values for Type-4 clones compared to the other related approaches. This depicts that utilizing TF-IDF as search technique and building meta-data with the help of identifiers works well in retrieving relevant clone methods. We notice a slight difference in recall performance between variants of Clone-Seeker with annotation and Clone-Seeker(Baseline). This means that there might still be room for improvement in terms of annotation strategies, to be investigated in the future. With 61% recall for semantic clones (WT3/T4), Clone-Seeker(Automatic) achieves the best performance score in the literature.

### 4.2. Code Clone Search Using Natural Language Query

In this section, we demonstrate the effectiveness of Clone-Seeker by applying natural language queries (natural-language-to-code search). These queries have been collected from Stack-Overflow[6]. Stack Overflow is a popular online programming community, where programmers ask questions about programming problems and give answers. The website has been found to be useful for software development (Diamantopoulos and Symeonidis, 2015; Ragkhitwetsagul et al., 2019) and

---

[6] https://stackoverflow.com/



Table 3: Recall scores(%) for Clone-Seeker and other related approaches on BigCloneBench

| Approaches | Clone Types | | | | | |
|---|---|---|---|---|---|---|
| | T1 | T2 | VST3 | ST3 | MT3 | WT3/T4 |
| *Clone search engines* | | | | | | |
| Clone-Seeker(Baseline) | 100 | 100 | 99 | 89 | 66 | 57 |
| Clone-Seeker(Manual) | 100 | 100 | 99 | 90 | 68 | 59 |
| Clone-Seeker(Automatic) | 100 | 100 | 100 | 90 | 68 | 61 |
| Siamese (Ragkhitwetsagul and Krinke, 2019) | 99 | 99 | 99 | 99 | 88 | 17 |
| FaCoY (Kim et al., 2018) | 65 | 90 | 67 | 69 | 41 | 10 |
| *Clone detectors (Sajnani et al., 2016)* | | | | | | |
| SourcererCC | 100 | 98 | 93 | 61 | 5 | 0 |
| CCFinderX | 100 | 93 | 62 | 15 | 1 | 0 |
| Deckard | 60 | 58 | 62 | 31 | 12 | 1 |
| iClone | 100 | 82 | 82 | 24 | 0 | 0 |
| NiCad | 100 | 100 | 100 | 95 | 1 | 0 |

also valuable for educational purposes (Nasehi et al., 2012). On Stack Overflow, each conversation contains a question and a list of answers. The answer frequently contains at least one code snippet as a solution to the question asked. Sometimes, the question itself also contains a code snippet. This usually indicates that a developer asks for either a more optimized solution than the one he posts or wishes to discuss some problem in the code. We demonstrate how clone methods are retrieved based on a natural language query with the help of an example (Table 5). Suppose that a developer is interested in searching for clone methods implementing the "Instantiate class object using reflection" functionality. The natural language query is first pre-processed and words are extracted as explained in Section 3.3. Then, this flat list of words is fed into Clone-Seeker(Manual), which generates top-3 most similar natural language documents. Manual annotation is highlighted with blue color in the natural language documents. These documents are mapped to their associated clone methods, which are finally presented to the user.

*Experiment Design.* The main aim of this experiment is to assess whether our methodology can help in retrieving clone methods using a natural language query. In this experiment, we compare the search results obtained from the three annotation strategies in Clone-Seeker: Baseline, Manual Annotation, and Automatic Annotation, as described in Section 3.3.

We build a benchmark of 43 queries belonging to the functionality types (clone classes) in BigCloneBench. We type in the keywords belonging to each clone class in the Stack Overflow website and search for the relevant post queries. The queries are manually selected based on two



requirements from Stack Overflow: (1) the associated post is related to "Java" and (2) the post includes a code snippet belonging to one of the mentioned functionality types in BigCloneBench. The full list of the 43 selected queries can be found in Table 4.

To evaluate search effectiveness, we feed Clone-Seeker with benchmark queries belonging to different clone classes after applying pre-processing steps as mentioned in Section 3.3 (*Word Extraction*). Then, we inspect the top-10 results retrieved from search corpus, as built from Big-CloneBench references and IJaDataset in Section 3.4 by identifying their clone classes. Afterwards, we calculate the precision, which is the number of ground-truth answers hit on average in the Top@k returned for a query using a Precision@k metric defined as follows:

$$Precision@k = \frac{1}{|Q|} \sum_{|Q|}^{i=1} \frac{|relevant_{i,k}|}{k} \qquad (3)$$

where $relevant_{i,k}$ represents the relevant search results for query i in the top k returned results, and Q is a set of queries. Precision shows the relevance of the returned results to the queries with respect to the ground-truth answers; the higher the value, the more relevant the results are. Moreover, we measure the Mean Reciprocal Rank (MRR) scores (Kim et al., 2018), which is the average of the reciprocal ranks of results of a set of queries Q. The reciprocal rank of a query is the inverse of the rank of the first hit result. It is used, when user is interested in considering only the first hit. MRR is calculated by using the following formula:

$$MRR = \frac{1}{|Q|} \sum_{|Q|}^{i=1} \frac{1}{rank_i} \qquad (4)$$

where $rank_i$ refers to the rank position of the first relevant result for the i-th test input method. The higher the MRR value is, the better the code search performs. Table 6 shows the evaluation results in terms of MRR and precision of different approaches for each query in the benchmark.

*Results.* The overall approach we propose leads to promising results in retrieving clone methods against specified natural language queries. Table 6 presents the results in terms of MRR, Precision@1, Precision@5, and Precision@10. Figure 2 presents a comparison between Clone-Seeker(Baseline), Clone-Seeker(Manual), and Clone-Seeker(Automatic) in terms of average MRR, Precision@1, Precision@5, and Precision@10.



Table 4: Benchmark Queries extracted from Stack-Overflow

| Id | Description |
|---|---|
| 2 | How to download and save a file from the Internet using Java? |
| 3 | How can I generate an MD5 hash? |
| 4 | Error while copying files from source to destination java |
| 5 | Java decompress archive file |
| 6 | Java: Accessing a File from an FTP Server |
| 7 | Basic Bubble Sort with ArrayList in Java |
| 8 | GEF editor functionality to view |
| 9 | ScrollingGraphicalViewer Select and Unselect listener |
| 10 | Java: Rollback Database updates? |
| 11 | Creating a Eclipse Java Project from another project, programatically |
| 12 | Java Display the Prime Factorization of a number |
| 13 | Random shuffling of an array |
| 14 | First occurrence in a binary search |
| 15 | Java - How to Load a Custom Font From a Resources Folder |
| 17 | Issues in RSA encryption in Java class |
| 18 | How can I play sound in Java? |
| 19 | Is there a way to take a screenshot using Java and save it to some sort of image? |
| 20 | printing the results of a fibonacci series |
| 21 | implementing GAE XMPP service as an external component to an existing XMPP server (e.g. ejabberd or OpenFire) |
| 22 | How to Encrypt/Decrypt text in a file in Java |
| 23 | Resize an Array while keeping current elements in Java? |
| 24 | How to open the default webbrowser using java |
| 25 | Open a file using Desktop(java.awt) |
| 26 | Java: get greatest common divisor |
| 27 | How to invoke a method in java using reflection |
| 28 | Parsing XML file with DOM (Java) |
| 29 | Change date format in a Java string |
| 30 | How to create a zip file in Java |
| 31 | Is it possible to select multiple directories at once with JFileChooser |
| 32 | How can I send an email by Java application using GMail, Yahoo, or Hotmail? |
| 33 | File containing its own checksum |
| 34 | Launching external process from Java : stdout and stderr |
| 35 | With Java reflection how to instantiate a new object, then call a method on it? |
| 36 | Create MySQL database from Java |
| 37 | Reading a binary input stream into a single byte array in Java |
| 38 | Get MAC Address of System in Java |
| 39 | How to delete a folder with files using Java |
| 40 | Fastest way to read a CSV file java |
| 41 | Transposing a matrix from a 2D array |
| 42 | Using Regular Expressions to Extract a Value in Java |
| 43 | How to copy an entire content from a directory to another in Java? |
| 44 | Check string for palindrome |
| 45 | java pdf file write |



Table 5: Example of retrieving relevant clone methods on the basis of natural language query

| Search query | With Java reflection how to instantiate a new object, then call a method on it? |
|---|---|
| **Search query (Word Extraction)** | with java reflect instanti new object call method |
| *Relevant Results (Natural Language Documents)* | |
| **Top 1** | get reflect class object use find constructor then instanti call arg con except runtim new instanc e1 |
| **Top 2** | get reflect class object use find constructor then instanti layout instanc string name list except method new |
| **Top 3** | get reflect class object use find constructor then instanti call arg throwabl ctor new instanc java lang invoc target except util wrap |
| *Clone Methods* | |
| **Top 1** | ```java
public Object callConstructor(Class c, Class[] classes, Object[] args) {
    Constructor con = null;
    try {
        con = c.getConstructor(classes);
    } catch (Exception e) {
        throw new RuntimeException("Error locating constructor: " + c);
    }
    try {
        return con.newInstance(args);
    } catch (Exception e1) {
        throw new RuntimeException("Error calling constructor: " + c);
    }
}
``` |
| **Top 2** | ```java
private static Layout getInstance(String className, Object[] list)
throws Exception {
    Constructor method = getConstructor(className);
    return (Layout) method.newInstance(list);
}
``` |
| **Top 3** | ```java
public Object call(Object[] args) throws Throwable {
    try {
        return ctor.getConstructor().newInstance(args);
    } catch (java.lang.reflect.InvocationTargetException e) {
        throw Utils.wrapInvocationException(e);
    }
}
``` |

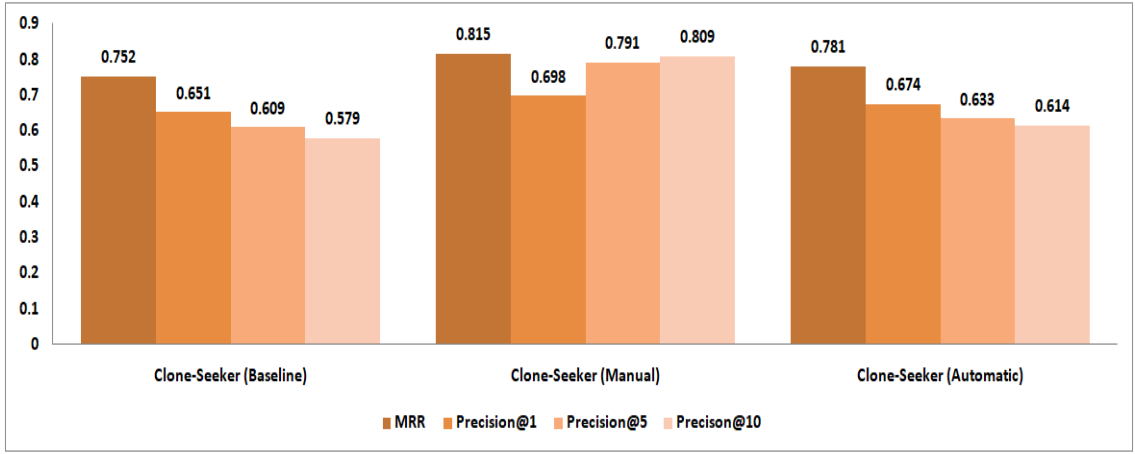

Figure 2: The Comparison Between Clone-Seeker(Baseline), Clone-Seeker(Manual), and Clone-Seeker(Automatic) in terms of average accuracy measures

Table 6: Evaluation results of Stack Overflow queries in terms of MRR(%) and precision (P@k)
('○' = Case 1, '◖' = Case 2, '◗' = Case 3, '*' = Case 4, '+' = Case 5)

| Id | Clone-Seeker(Baseline) | | | | Clone-Seeker(Manual) | | | | Clone-Seeker(Automatic) | | | |
|---|---|---|---|---|---|---|---|---|---|---|---|---|
| | MRR | P@1 | P@5 | P@10 | MRR | P@1 | P@5 | P@10 | MRR | P@1 | P@5 | P@10 |
| **2**○ | 0.167 | 0 | 0 | 0.1 | 0.5 | 0 | 0.6 | 0.6 | 0.2 | 0 | 0.2 | 0.1 |
| **3**○ | 1 | 1 | 0.8 | 0.9 | 1 | 1 | 1 | 1 | 1 | 1 | 0.8 | 0.9 |
| **4*** | 1 | 1 | 1 | 1 | 1 | 1 | 1 | 1 | 1 | 1 | 1 | 1 |
| **5**○ | 0.33 | 0 | 0.2 | 0.1 | 0.5 | 0 | 0.4 | 0.6 | 0.33 | 0 | 0.2 | 0.1 |
| **6*** | 1 | 1 | 1 | 1 | 1 | 1 | 1 | 1 | 1 | 1 | 1 | 1 |
| **7*** | 1 | 1 | 1 | 1 | 1 | 1 | 1 | 1 | 1 | 1 | 1 | 1 |
| **8**◗ | 0.25 | 0 | 0.2 | 0.2 | 0.33 | 0 | 0.2 | 0.2 | 0.33 | 0 | 0.2 | 0.2 |
| **9**○ | 1 | 1 | 0.6 | 0.3 | 1 | 1 | 0.6 | 0.4 | 1 | 1 | 0.6 | 0.3 |
| **10*** | 1 | 1 | 1 | 1 | 1 | 1 | 1 | 1 | 1 | 1 | 1 | 1 |
| **11**○ | 0.5 | 0 | 0.4 | 0.5 | 0.5 | 0 | 0.8 | 0.9 | 0.5 | 0 | 0.4 | 0.5 |
| **12*** | 1 | 1 | 1 | 1 | 1 | 1 | 1 | 1 | 1 | 1 | 1 | 1 |
| **13*** | 1 | 1 | 1 | 1 | 1 | 1 | 1 | 1 | 1 | 1 | 1 | 1 |
| **14**+ | 1 | 1 | 0.8 | 0.9 | 0.5 | 0 | 0.8 | 0.9 | 1 | 1 | 0.8 | 0.9 |
| **15**○ | 1 | 1 | 0.4 | 0.2 | 1 | 1 | 1 | 1 | 1 | 1 | 0.4 | 0.2 |
| **17**○ | 0.1 | 0 | 0 | 0.1 | 0.167 | 0 | 0 | 0.2 | 0.1 | 0 | 0 | 0.1 |
| **18*** | 1 | 1 | 1 | 1 | 1 | 1 | 1 | 1 | 1 | 1 | 1 | 1 |
| **19**◗ | 1 | 1 | 0.8 | 0.8 | 1 | 1 | 1 | 1 | 1 | 1 | 1 | 1 |
| **20**◖ | 1 | 1 | 1 | 0.5 | 0.5 | 0 | 0.8 | 0.9 | 1 | 1 | 1 | 1 |
| **21**○ | 1 | 1 | 0.6 | 0.6 | 1 | 1 | 0.8 | 0.7 | 1 | 1 | 0.6 | 0.7 |
| **22*** | 0.11 | 0 | 0 | 0.2 | 0.143 | 0 | 0 | 0.2 | 0.11 | 0 | 0 | 0.2 |
| **23**+ | 1 | 1 | 1 | 1 | 1 | 1 | 1 | 0.9 | 1 | 1 | 1 | 1 |
| **24**○ | 1 | 1 | 0.4 | 0.3 | 1 | 1 | 0.6 | 0.7 | 1 | 1 | 0.4 | 0.3 |
| **25**◖ | 0.5 | 0 | 0.4 | 0.3 | 0.33 | 0 | 0.4 | 0.6 | 1 | 1 | 0.6 | 0.4 |
| **26**○ | 1 | 1 | 0.2 | 0.2 | 1 | 1 | 1 | 1 | 1 | 1 | 0.2 | 0.3 |
| **27**○ | 0.5 | 0 | 0.8 | 0.6 | 1 | 1 | 1 | 0.8 | 0.5 | 0 | 0.8 | 0.6 |
| **28**○ | 1 | 1 | 1 | 0.9 | 1 | 1 | 1 | 1 | 1 | 1 | 1 | 0.9 |
| **29**◖ | 0.2 | 0 | 0.2 | 0.4 | 1 | 1 | 0.4 | 0.3 | 1 | 1 | 0.6 | 0.6 |
| **30*** | 0.33 | 0 | 0.6 | 0.8 | 0.5 | 0 | 0.6 | 0.8 | 0.5 | 0 | 0.6 | 0.8 |
| **31**○ | 1 | 1 | 0.2 | 0.2 | 1 | 1 | 0.8 | 0.8 | 1 | 1 | 0.2 | 0.5 |
| **32**○ | 1 | 1 | 0.6 | 0.6 | 1 | 1 | 1 | 1 | 1 | 1 | 0.6 | 0.6 |
| **33**○ | 1 | 1 | 0.6 | 0.7 | 1 | 1 | 1 | 1 | 0.5 | 0 | 0.6 | 0.7 |
| **34**○ | 0.5 | 0 | 0.8 | 0.6 | 1 | 1 | 1 | 0.9 | 0.5 | 0 | 0.8 | 0.6 |
| **35**○ | 1 | 1 | 0.4 | 0.2 | 1 | 1 | 0.8 | 0.8 | 1 | 1 | 0.6 | 0.4 |
| **36**○ | 0 | 0 | 0 | 0 | 1 | 1 | 0.6 | 0.7 | 0.143 | 0 | 0 | 0.1 |
| **37**○ | 1 | 1 | 0.4 | 0.2 | 1 | 1 | 1 | 1 | 1 | 1 | 0.4 | 0.2 |
| **38*** | 1 | 1 | 1 | 0.8 | 1 | 1 | 1 | 0.8 | 1 | 1 | 1 | 0.8 |
| **39**○ | 1 | 1 | 1 | 0.8 | 1 | 1 | 1 | 1 | 1 | 1 | 0.8 | 0.6 |
| **40**○ | 0.25 | 0 | 0.2 | 0.2 | 0.333 | 0 | 0.6 | 0.5 | 0.25 | 0 | 0.2 | 0.2 |
| **41*** | 1 | 1 | 1 | 1 | 1 | 1 | 1 | 1 | 1 | 1 | 1 | 1 |
| **42**○ | 1 | 1 | 0.8 | 0.8 | 1 | 1 | 1 | 1 | 1 | 1 | 0.8 | 0.8 |
| **43**○ | 0.11 | 0 | 0 | 0.2 | 0.25 | 0 | 0.4 | 0.7 | 0.11 | 0 | 0 | 0.2 |
| **44*** | 1 | 1 | 1 | 1 | 1 | 1 | 1 | 1 | 1 | 1 | 1 | 1 |
| **45**○ | 0.5 | 0 | 0.8 | 0.7 | 0.5 | 0 | 0.8 | 0.9 | 0.5 | 0 | 0.8 | 0.6 |
| **Average** | **0.752** | **0.651** | **0.609** | **0.579** | **0.815** | **0.698** | **0.791** | **0.809** | **0.781** | **0.674** | **0.633** | **0.614** |

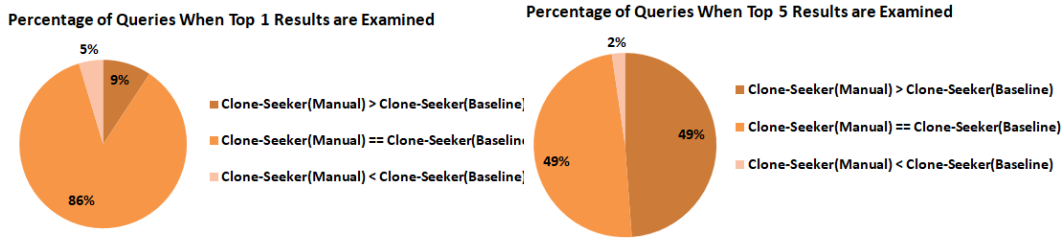

Figure 3: The Percentage of Queries that Clone-Seeker(Manual) performs Better/Worse than Baseline

We notice from Table 6 and Figure 2 that Clone-Seeker(Automatic) obtains an average MRR score of 0.781, which outperforms the Clone-Seeker(Baseline) (0.752) by 3.86%. Clone-Seeker(Automatic) achieves 3.53%, 3.94%, and 6.04% improvement with respect to Clone-Seeker(Baseline) in terms of average Precision@1, Precision@5, and Precision@10 respectively. Moreover, Clone-Seeker(Manual) achieves better performance than Clone-Seeker(Automatic) in terms of MRR and precision. Clone-Seeker(Manual), in turn, works well as compared to Clone-Seeker(Baseline) in terms of MRR and precision. In general, the Clone-Seeker(Manual) approach performs the best for the selected benchmark queries in retrieving relevant clone methods.

We further compare the performance of different variants of Clone-Seeker in terms of queries, as followed by Lv et al. (2015). Figure 4 displays the percentage of queries that Clone-Seeker(Automatic) performs better/worse than the Clone-Seeker(Baseline) approach. We can see that when the top 1 returned results are investigated, Clone-Seeker(Automatic) outperforms in 5% of the queries, equals in 93% and underperforms in none. In terms of the top 5 results, Clone-Seeker(Automatic) performs better than the Clone-Seeker(Baseline) approach in 12% queries, equals in 86% and underperforms in only 2% of the queries. The results confirm that the improvement achieved by Clone-Seeker(Automatic) is significant in retrieving relevant clone methods. Moreover, we notice from Figure 5, that the improvement achieved by Clone-Seeker(Manual) is better as compared to Clone-Seeker(Automatic) in terms of retrieving relevant clone methods, when top-1 and top-5 results have been examined. Finally, Figure 3 depicts that Clone-Seeker(Manual) outperforms Clone-Seeker(Baseline) in terms of retrieving relevant clone methods, when top 1 and top 5 results have been investigated.

In our qualitative investigation, we experience five different scenarios, represented with special symbols involving the precision values (Table 6), with benchmark queries (Table 4) being fed into Clone-Seeker using different annotation strategies:

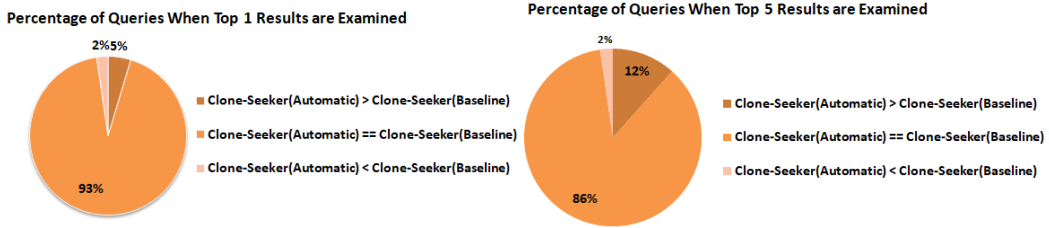

Figure 4: The Percentage of Queries that Clone-Seeker(Automatic) performs Better/Worse than Baseline

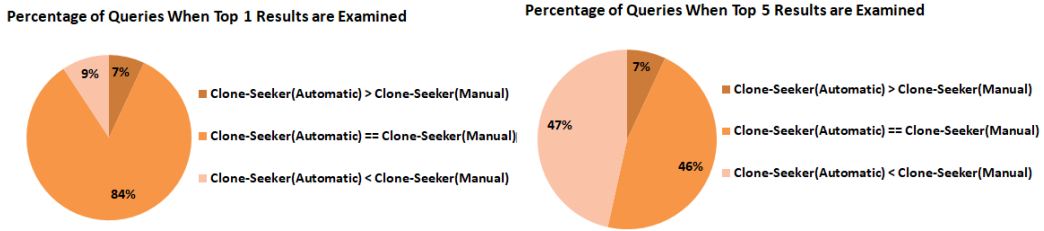

Figure 5: The Percentage of Queries that Clone-Seeker(Automatic) Performs Better/Worse than Clone-Seeker(Manual)

1. Case 1(∘): The first scenario is when Clone-Seeker(Manual) performs better than Clone-Seeker(Automatic) and Clone-Seeker(Baseline) in retrieving clone methods against specified queries (found in 24 cases). This means that the specified queries achieve high TF-IDF scores when comparing with the natural language document of clone methods generated with Clone-Seeker(Manual) approach as compared to other approaches.

2. Case 2(◁): The second scenario is when Clone-Seeker(Automatic) works better as compared to Clone-Seeker(Manual) and Clone-Seeker(Baseline) in retrieving clone methods against specified queries (found in 3 cases). This depicts that the specified queries accomplish high TF-IDF score, when comparing with the natural language document of clone methods produced with Clone-Seeker(Automatic) approach as compared to other approaches.

3. Case 3(▷): The third scenario is when Clone-Seeker(Manual) and Clone-Seeker(Automatic) both perform better than Clone-Seeker(Baseline) in retrieving clone methods against specified queries (found in 2 cases). This illustrates that the specified queries attain high TF-IDF score, when comparing with the natural language document of clone methods generated with Clone-Seeker(Manual) and Clone-Seeker(Automatic) approaches as compared to baseline.

4. Case 4(*): The fourth scenario is when baseline, manual and automatic annotations equally performs well in retrieving clone methods against specified queries (found in 12 cases). This means that the specified queries achieve equal TF-IDF scores, when comparing with the natural language document of clone methods produced with all the approaches such as Clone-Seeker(Manual), Clone-Seeker(Automatic), and Clone-Seeker(Baseline).

5. Case 5(+): The fifth scenario is when Clone-Seeker(Baseline) and Clone-Seeker(Automatic) performs well as compared to Clone-Seeker(Manual) in retrieving clone methods against specified queries(found in 2 cases). This means that the specified queries attain high TF-IDF scores, when comparing with the natural language document of clone methods generated with Clone-Seeker(Baseline) and Clone-Seeker(Automatic) approaches as compared to Clone-Seeker(Manual) approach.

We do not find any scenario when Clone-Seeker(Baseline) outperforms both Clone-Seeker(Manual) and Clone-Seeker(Automatic) in retrieving clone methods against specified queries, which indicates that the annotation strategy works quite well in performing effective code clone search.

## 5. Limitations

There are certain limitations of this work, which can be further explored and mitigated in the future. We apply a threshold value of 10, based on a limited preliminary study on a small random subset, to retrieve the most recurrent keywords from each clone class. Different threshold values can be applied to get more improved results. Clone-Seeker also depends upon how a user represent each clone class manually. A more accurate manually annotated description can be applied to get more effective results. Some limitations originate from the selected dataset. While the dataset for this study is based on well-known code clone dataset (i.e. BigCloneBench), it does not necessarily mean the codebase represents Java language source code entirely. This is a a threat to external validity in terms of generalizability, and further experiments are needed to claim the applicability of our approach for any Java code. We only apply word counting technique to extract the keywords automatically. In the future, we plan to investigate several keywords extraction techniques such as word degree, TF-IDF (which we use for searching in this work, not for keyword extraction), and RAKE (Nasar et al., 2019).

We have used 43 queries to validate the effectiveness of Clone-Seeker approaches for natural language query. Although these queries are real-world queries collected from the Stack-Overflow website, admittedly they do not cover all types of queries that a developer may ask. Also, although our codebase consists of 8 million clone pairs, it is just a tiny sample of all available source code.

In the future, we plan to reduce the threats to external threats by investigating more queries over a much larger codebase.

## 6. Conclusion and Future Work

In this paper, we propose a novel approach of retrieving clone methods effectively. Our approach can retrieve clone methods effectively, based on a search query in terms of source code terms as well as natural language. We apply different annotation strategies to build metadata (i.e. natural language document) for clone methods . We successfully demonstrate the effectiveness of Clone-Seeker approaches through empirical evaluation. We achieve the best recall for Type-4 clones as compared to the state-of-the art. Similarly, we demonstrate the effectiveness of our approach by applying natural language queries collected from Stack-Overflow posts.

While we have promising results, our work can be extended in various directions. In this study, our focus is on introducing an effective technique for retrieving clone methods. However, code clones can be of different types and granularity levels (e.g. simple clones, structural clones, file clones, and clones of other artifact types such as models (Hammad et al., 2020b; Babur et al., 2019)). We plan to investigate whether our approach can be used for different clone types and granularities in the future. We only use the TF-IDF technique in searching and retrieval as a proof of concept, however there are several other information retrieval techniques such as word2vec (Mikolov et al., 2013), glove (Pennington et al., 2014), etc. , which can be applied and comparatively evaluated. Similarly, neural networks can also be applied to build an effective search approach. Joint neural network models (Gu et al., 2018) for the natural language documents and clone methods can be explored to have effective search results. In theory, our approach can search for clone methods written in any programming language. This can be done by first detecting method-level code clones using some clone detector tool such as SourcererCC (Sajnani et al., 2016), NiCAD (Cordy and Roy, 2011) and Clone Miner (Basit and Jarzabek, 2009) to get clone references. Clone detection can be performed by following a similar process as used for BigCloneBench, by applying heuristics along with manual validation. Similarly, if comments are a reliable source of information in some particular dataset, then they can also be utilized in annotating clone classes. This may help in effectively retrieving clone methods based on search query.

## 7. Acknowledgment

We acknowledge SURFsara for providing us computational credits for the experiments.